\documentclass[journal=jctcce,manuscript=article]{achemso}
\usepackage{graphics,graphicx,amsfonts,amsmath,amsbsy,amssymb,color}
\usepackage{verbatim}  
\usepackage{psfrag}  
\usepackage{tabularx}
\usepackage{xmpmulti}
\usepackage{hyperref}
\usepackage{marginnote}
\usepackage{subfigure}
\usepackage{paracol}
\usepackage{xparse}
\usepackage{layouts}
\usepackage{afterpage}
\usepackage{braket}
\usepackage{paralist}
\usepackage{bm}
\usepackage{url}
\usepackage[normalem]{ulem}
\usepackage{color}
\usepackage[colorinlistoftodos]{todonotes}
\usepackage[american]{babel} 
\usepackage[version=3]{mhchem}

\newcommand{\refeq}[1]{{Eq.~(\ref{#1})}}
\newcommand{\reffig}[1]{{Fig.~\ref{#1}}}

\newcommand{\iFCIQMC}{\ensuremath{i-}FCIQMC\ }
\newcommand{\iFCIQMCbracket}{\ensuremath{i-}FCIQMC}
\newcommand{\bz}[1]{{{\ce{C6H6\bond{-}#1}}}}

\title{Fully quantum embedding with density functional theory for full configuration interaction quantum Monte Carlo}

\author{Hayley~R.~Petras}
\altaffiliation{These authors contributed equally to this paper}
\affiliation[First University]
{Department of Chemistry, University of Iowa}
\alsoaffiliation[Another]
{University of Iowa Informatics Initiative, University of Iowa}

\author{Daniel~S.~Graham}
\altaffiliation{These authors contributed equally to this paper}
\affiliation[Second University]
{Department of Chemistry, University of Minnesota}

\author{Sai~Kumar~Ramadugu}
\affiliation[First University]
{Department of Chemistry, University of Iowa}
\alsoaffiliation[Another]
{University of Iowa Informatics Initiative, University of Iowa}

\author{Jason~D.~Goodpaster}
\affiliation[Second University]
{Department of Chemistry, University of Minnesota}

\author{James~J.~Shepherd}
\email{james-shepherd@uiowa.edu}
\affiliation[First University]
{Department of Chemistry, University of Iowa}
\alsoaffiliation[Another]
{University of Iowa Informatics Initiative, University of Iowa}
\date{\today}

\begin{document}

\begin{abstract}

We here develop a fully-quantum embedded version of initiator full configuration interaction quantum Monte Carlo (\iFCIQMCbracket) and apply it to study an ionic bond (lithium hydride, LiH) and a covalent bond (hydrogen flouride, HF) physisorbed to a benzene molecule.
The embedding is performed using a recently-developed Huzinaga projection operator approach, which affords good synergy with \iFCIQMC by minimizing the number of orbitals in the calculation.
When considering the dissociation energy of these bonds into closed-shell ionic fragments, we find that \iFCIQMCbracket embedded in density functional theory (\iFCIQMCbracket-in-DFT) delivers comparable accuracy with coupled cluster singles and doubles with perturbative triples embedded in density functional theory (CCSD(T)-in-DFT).
In treating the bond dissociation energy curve of (HF) \iFCIQMCbracket-in-DFT has improved accuracy over CCSD(T)-in-DFT due to the presence of strong correlation. 
We discuss the implications of the new \iFCIQMCbracket-in-DFT method as applied to bond breaking in catalysis. 

\end{abstract}

\section{Introduction} %

Catalysis often involves bond rearrangements at surfaces, a process featuring closely-separated energy minima, stretched bonds, and transition states.
The electronic structure of these systems can become extremely complex; combined with energy differences that can be sub-millihartree, systematic study of catalytic bond rearrangements necessitates the development of new high-accuracy quantum chemistry methods. Although this is a subject of active and ongoing investigation, the high cost of wavefunction methods in particular prevents their widespread application.

One such method is full configuration interaction quantum Monte Carlo (FCIQMC) and its initiator adaptation (\iFCIQMCbracket), which are both members of a family of particularly attractive high-accuracy electronic structure methods that seek to combine the exactness of full configuration interaction (FCI) with the increased speed achieved by quantum Monte Carlo (QMC).\cite{booth_fermion_2009,cleland_initiator_2009} The first FCIQMC paper showed that the FCI ground-state wavefunction could be stochastically sampled due to the sparsity in the Hamiltonian; \cite{booth_fermion_2009} it had previously been considered that there was no way to sample such a large vector as the exact FCI wavefunction. Since this pioneering work, many adaptions to FCIQMC and \iFCIQMC have been developed successfully for calculating correlation energies of a wide variety of benchmark systems. 

\iFCIQMC has already been used for a variety of applications on relatively small systems, including model systems (such as the Hubbard model\cite{schwarz_insights_2015,spencer_sign_2012} and the uniform electron gas\cite{shepherd_investigation_2012,shepherd_full_2012}) and dimers (such as \ce{C2}\cite{booth_breaking_2011} and \ce{Cr2} \cite{booth_linear-scaling_2014}).  It has also seen real applications that are more ambitious, such as iron porphyrins, which used a complete active space adaptation,\cite{li_manni_combining_2016}  and fully periodic nickel oxide chains \cite{booth_towards_2013}. 
A significant amount of investigation has also been aimed at using the full scheme and initiator adaption of FCIQMC to stochastically sample reduced density matrices within the FCIQMC method. \cite{overy_unbiased_2014,booth_explicitly_2012,thomas_analytic_2015,thomas_stochastic_2015, blunt_density_2017,blunt_semi-stochastic_2015}
Altogether, FCIQMC and its adaptations seem well-poised for answering important questions about the electronic structures of complex chemical systems with high accuracy. 

Unfortunately, like all of its high-accuracy cousin methods, \iFCIQMC is limited in its scope by its high cost: it can only treat relatively small system sizes (which here means number of electrons). Many further adaptations to \iFCIQMC have been developed to allow for the application of \iFCIQMC to larger systems.  These adaptations include a combination of complete active space self-consistent field (CASSCF) with \iFCIQMC \cite{li_manni_combining_2016}, the semi-stochastic projector Monte Carlo method \cite{petruzielo_semistochastic_2012}, model space QMC,\cite{ten-no_stochastic_2013} \cite{ohtsuka_study_2015} heat-bath configuration interaction,\cite{holmes_heat-bath_2016} perturbation theory \cite{blunt_communication:_2018}, stochastic multi-configurational self-consistent field theory (MCSCF) utilizing the FCIQMC methodology \cite{thomas_stochastic_2015}, use of a transcorrelated Hamiltonian with \iFCIQMC \cite{sharma_spectroscopic_2014,luo_combining_2018}, and combinations of the above methods, such as semistochastic heat-bath CI \cite{sharma_semistochastic_2017,holmes_excited_2017,li_fast_2018,chien_excited_2018}.

These efforts are made all the more relevant because there are also varieties of FCIQMC which broaden its applicability. Density matrix QMC,\cite{blunt_density-matrix_2014} has been developed for temperature-dependent electronic structure. Additionally, several FCIQMC methods have been developed for use on excited states, such as changing the underlying propagator \cite{booth_communication:_2012}, the Krylov-projected QMC method \cite{blunt_krylov-projected_2015}, utilizing a L\"{o}wdin partitioning technique\cite{ten-no_stochastic_2013}, using a Gram--Schmidt procedure \cite{blunt_excited-state_2015}, and by restricting the population to the orthogonal complement of the low lying states \cite{humeniuk_excited_2014}. The stochastic approach in the Slater determinant space has also been studied on the coupled cluster equations, called coupled cluster Monte Carlo. \cite{thom_stochastic_2010,franklin_linked_2016,deustua_converging_2017,scott_stochastic_2017}  FCIQMC has also been adapted to treat the Clock Hamiltonian, to simulate the full time evolution of a quantum system. \cite{mcclean_clock_2015} A deterministic version of FCIQMC has been developed, \cite{tubman_deterministic_2016} as well as a fast randomized iteration framework to essentially perform FCIQMC without walkers. \cite{greene_beyond_2019} We also note that there are a number of methods which fall under the umbrella of selected configuration interaction (CI), where the CI is solved deterministically, which form a distinct and related family of methods.\cite{scemama_excitation_2018, garniron_selected_2018, dash_perturbatively_2018}

Quantum embedding methods were specifically developed to reduce the problem of scaling present in high level-methods such as \iFCIQMCbracket. Embedding methods limit high-level calculations to a small subsystem that is embedded in the potential arising from the rest of the system, reducing the overall computational cost. When highly accurate embedding potentials are used, good accuracy can be achieved even when a subsystem is limited to a few atoms; therefore, embedding methodologies have been  successfully applied to a wide variety of systems. \cite{French2001FromEmbedding, French2003AssignmentEmbedding, Chung2015TheApplications, Vreven2003InvestigationMethod, Joshi2005EmbeddedHZSM-5, Sokol2004HybridMaterials,pavanello_modelling_2011, pavanello_subsystem_2013, doi:10.1021/acs.jctc.8b01112, doi:10.1021/acs.jctc.7b00666, doi:10.1021/acs.jctc.6b01065,doi:10.1021/acs.jctc.6b00685, doi:10.1021/ct5011032, doi:10.1021/acs.jctc.5b00630, doi:10.1021/jp511275e, doi:10.1063/1.5055942} Additionally, a large amount of work has been performed developing accurate embedding frameworks including quantum mechanics / molecular mechanics  (QM/MM),\cite{warshel1976theoretical} ONIOM,\cite{Svensson1996} density matrix embedding theory (DMET),\cite{Knizia2013} Green's function embedding,\cite{Onida, Chibani2016} and density functional theory (DFT) embedding.\cite{Jacob2014, Wesolowski2013, Wesoowski2006One-ElectronSystems, Neugebauer2010, Yang1991, Huang2011, Goodpaster2010ExactTheory} A recent review has considered the comparisons between DMET, Green's function embedding, and DFT embedding and we direct the interested reader to Ref.~\citenum{doi:10.1021/acs.accounts.6b00356}. 
Many wavefunction methods such as density-matrix renormalization group  (DMRG),\cite{doi:10.1021/acs.jctc.6b00476} coupled cluster singles and doubles with perturbative triples (CCSD(T)),\cite{manby2012simple} second order M\o ller-Plesset perturbation theory (MP2),\cite{doi:10.1021/acs.jctc.8b01112} and multireference configuration interaction (MRCI)\cite{doi:10.1063/1.5050533} have been embedded as the high-level theory; this work presents the first use of \iFCIQMC embedding.

The quantum embedding for this work was done using projection-based embedding,\cite{manby2012simple} which is DFT embedding method. Projection-based embedding is one solution to the non-additive kinetic energy problem of DFT embedding.\cite{Goodpaster2014AccurateWavefunctions, Lee2019Projection-BasedEmbedding} The initial projection operator applied to this problem was the $\mu$ projection operator developed by the Manby and Miller groups.\cite{manby2012simple} This projection operator allows two embedded DFT subsystems (DFT-in-DFT) to exactly recreate full-system Kohn-Sham DFT. However, when embedding a wavefunction (WF) subsystem within a DFT environment (WF-in-DFT), the number of orbitals in the WF subsystem is the same as the number of orbitals in the full system. Since WF methods scale poorly with number of orbitals, basis set truncation methods were developed to reduce computational cost.\cite{Barnes2013AccurateEmbedding, Bennie2015} The more recent truncation method removes basis functions from a subsystem when the density of that subsystem is below a threshold---a manner that maintains a high degree of accuracy. By decoupling the WF calculation from the total size of the system, WF-level energies may be calculated for systems consisting of hundreds of atoms.  The $\mu$ operator method has shown a high degree of accuracy for transition-metal and enzyme catalysis, and oxidation potentials of molecules in solution, among other systems of interest.\cite{Lee2019Projection-BasedEmbedding, Goodpaster2012DensityComplexes} Additionally, several groups have used the $\mu$ projection operator to embed multireference wavefunction methods for application to transition metal catalysts. \cite{Chapovetsky2018PendantReduction, deLimaBatista2017PhotophysicalApproach} These systems are inherently multireference; however, as the multireference character is localized to the metal center, $\mu$ embedding calculations were able to closely match experimental results. 

K\'{a}llay and co-workers introduced the Huzinaga projection operator for DFT embedding;\cite{jcp_145_64107} however, that work truncated the orbitals by using local correlation methods.  We showed that the Huzinaga projection operator could be used for aggressive truncation of the orbital space, where the densities could be absolutely localized on the atomic basis functions centered on atoms within the subsystem.\cite{Chulhai2017} This allows for high computational efficiency as the WF subsystem has a greatly reduced number of molecular orbitals. Huzinaga projection embedding has also been successfully extended to periodic systems,\cite{Chulhai2018Projection-BasedSystems} allowing for cluster or periodic WF calculations embedded in a periodic DFT environment. Given that the absolutely localized basis used in Huzinaga projection-based embedding reduces the number of orbitals to only those centered on the atoms of interest, we here determine the effectiveness of \iFCIQMC on a absolutely localized subsystem within the embedding potential of the full system.

We are generally motivated to increase the range and scope of systems available for study with \iFCIQMCbracket.
With a view toward our long-term interests in the study of bond-breaking and bond rearrangement on surfaces relevant to heterogeneous catalysis, we here study bond dissociation for diatomic molecules containing ionic or covalent bonds (specifically, LiH and HF, respectively) physisorbed onto a benzene molecule using \iFCIQMCbracket.
This type of calculation (with $\sim 35$ active electrons) is currently at the edge of applicability for \iFCIQMCbracket; sometimes the system can be treated, and other times it cannot be treated.
We show that embedding greatly alleviates the cost scaling of our model system. 
Specifically, data show that \iFCIQMC calculations performed on the full system (including both the diatomic molecule and the benzene molecule) fails to converge, whereas the system in which the benzene is represented by embedding converges with the same efficiency as an isolated molecular calculation. 
We analyze the type of convergence behaviors in \iFCIQMC and relate them to the differing electronic structures of the dissociation reactants and products. In addition, we explore the applicability of \iFCIQMC to a range of atomic separations of HF on benzene by calculating a dissociation curve using both \iFCIQMC and CCSD(T). We show that for HF on benzene, \iFCIQMC does not have the same failure CCSD(T) shows in regions of strong correlation.

\section{Methods} 

\subsection{\iFCIQMC} %

Full configuration interaction quantum Monte Carlo \cite{booth_fermion_2009} and its initiator adaptation \cite{cleland_initiator_2009} attempt to solve for the ground-state wavefunction $|\Psi_0\rangle$ of the imaginary-time Schr\"{o}dinger equation of a given Hamiltonian $\hat{H}$: 
\begin{equation}
\frac{d|\Psi_0\rangle}{d\tau} = -\hat{H}|\Psi_0\rangle
\label{imagtSE}
\end{equation}
where $\tau$ represents imaginary time. 
Beginning with a wavefunction that has non-zero overlap with the ground state, this equation can be solved in the long-imaginary-time limit to give the ground state wavefunction:
\begin{equation}
     \lim_{\tau \to \infty} e^{-(\tau\hat{H}-S)}|D_0\rangle \propto |\Psi_0\rangle 
\end{equation}
where $|D_0\rangle$ is the reference Slater determinant, here taken to be the Hartree--Fock wavefunction. 
This relationship holds for any constant energy shift $S$.
When long enough imaginary time $\tau$ has passed, $S$ can be averaged, and the correlation energy ($E_\mathrm{corr}=E_\mathrm{total}-E_\mathrm{Hartree-Fock}$) found.

The full configuration interaction wavefunction is typically written as a sum of Slater determinants, $|D_i\rangle$, 
\begin{equation}
|\Psi_0 \rangle = \sum_i {c_i |D_i\rangle}
\label{sumDi}
\end{equation}
As such, the imaginary time evolution operator acts in a determinant space.

Substituting \refeq{sumDi} into \refeq{imagtSE} gives an expression which can be written as a finite difference
\begin{equation}
    c_i^{m+1} - c_i^m = c_i^m \tau ( -H_{ii}+S )-
    \sum_{j\neq i} {c_j^m \tau H_{ij}}.
\end{equation}
Here, $c_i^{m}$ is the coefficient of the $i^\mathrm{th}$ determinant at the $m^\mathrm{th}$ iteration of the simulation (after which $m\tau$ units of imaginary time have elapsed). 
The Hamiltonian is represented in the Slater determinant basis as:
\begin{equation}
    H_{ij} = \langle D_i|\hat{H}|D_j\rangle .
\end{equation} 
In the original FCIQMC algorithm, the weight $c_i$ takes integer values. \cite{booth_fermion_2009}
The walker population $N_w$ is given by $N_w=\sum_i c_i$. When $S$ is varied to keep the walker population constant, its average becomes an estimate of the total ground-state energy.

The population of particles evolves towards the ground state using the following three steps introduced by Booth et al: \begin{enumerate} 
\item The particles with weight $c_{i}$ are allowed to spawn from site $i$ to a connected site $j$, where $H_{ij} \neq 0$ and $i \neq j$. The probability of spawning, $p(j|i)$ is uniform over the $j$ which are connected by one or two electron excitations to $i$. 
The integer part of $\frac{H_{ij}\tau}{  p(j|i)}$ (including its sign) is then added to the weight at $j$. The non-integer remainder $r$ is added with probability $|r|$ as  $\pm1$, where the sign comes from the sign of $r$. 
\item Each particle with weight $c_i$ changes its weight by $|S-H_{ii}|\tau$. As above, the integer part of $|S-H_{ii}|\tau$ is added to the weight at $i$. The non-integer remainder $r$ treated as above.
\item Pairs of particles on the same site with opposite weight $c_i$ annihilate each other and and are removed from the simulation, leaving a population containing only a single sign on each site. \end{enumerate} 
FCIQMC is not restricted to using only integer weights $c_i$.  Real weights can be used; this adds a step to the above algorithm where the real weight is rounded off stochastically below a certain threshhold (here, 0.01), chosen to reduce stochastic error and raise efficiency \cite{petruzielo_semistochastic_2012}.

The initiator adaption to FCIQMC, \iFCIQMC, separates the Slater determinant space into those with $n_{add}$ (here, 3) or more walkers and those with fewer. If the origin of a spawning event (item 1. in the list above above) is not an ``initiator'' and the spawning is attempted onto a site without walkers, $H_{ij}$ is zeroed. The result is a dynamically-modified Hamiltonian, which profoundly influences convergence of the simulation. 
A simulation is only converged in the limit when changing the walker population no longer changes the energy (i.e., $N_w\rightarrow \infty$).
This is an important practical limitation that must be contended with when running an \iFCIQMCbracket calculation, and this ensures the wavefunction must be sampled with sufficient detail in order to attain statistical and systematic convergence. As $N_w \rightarrow \infty$, the full configuration interaction (i.e. exact) limit is achieved; away from this limit, the calculation contains a small error termed the initiator error. 
This error typically converges as $\sim\exp (-\alpha N_w)$ and is challenging to extrapolate away. 
Reducing this error is crucial to the success of \iFCIQMC; its pre-factor/rate of decay is highly system dependent, and for larger systems can bottleneck the calculations.

\subsection{Embedding} %
To perform \iFCIQMCbracket-in-DFT embedding, the full system density is first split into two subsystems, subsystem A and subsystem B
\begin{equation}
   \gamma^\text{tot} = \gamma^\text{A} + \gamma^\text{B}
\end{equation} where $\gamma^\text{A}$ and $\gamma^\text{B}$ are the densities matrices of subsystems A and B, respectively.  We then obtain the DFT densities of the subsystems  through a freeze-and-thaw algorithm.\cite{Chulhai2017} This algorithm works by iteratively relaxing the density of subsystem A within the embedding potential and projection operator generated by the frozen density of subsystem B, and then freezing the subsystem A density and relaxing the subsystem B density within the embedding potential and projection operator generated by subsystem A until both subsystem densities have converged.  The Fock matrix of subsystem A embedded in  subsystem B can be written as
\begin{equation}
	\mathbf{F}^\text{A-in-B} = \mathbf{h}^\text{A-in-B}[\gamma^\text{A},\gamma^\text{B}] + \mathbf{g}[\gamma^\text{A}]
\end{equation}
where $\mathbf{g}$ contains the Coulomb and exchange-correlation potential for DFT---and the embedded core Hamiltonian is

\begin{equation}
	\mathbf{h}^\text{A-in-B}[\gamma^\text{A}, \gamma^\text{B}] = \mathbf{h} + \mathbf{g}[\gamma^\text{A}+\gamma^\text{B}] - \mathbf{g}[\gamma^\text{A}] + \mathbf{P}^\text{B}
    \label{eq:mol-embedded-core}
\end{equation}

\noindent where $\mathbf{h}$ is the
one electron Hamiltonian, and thus contains the kinetic and nuclear potential operators for both subsystems, and $\mathbf{P}^\text{B}$ is the Huzinaga projection operator for subsystem A, given by
\begin{equation}
   \mathbf{P}^\text{B} = - \frac{1}{2} \left( \mathbf{F}^\text{AB} \gamma^\text{B} \textbf{S}^\text{BA} + \textbf{S}^\text{AB} \gamma^\text{B} \mathbf{F}^\text{BA} \right),
   \label{eq:huzinaga-operator}
\end{equation} where $\mathbf{F}^\text{AB}$ and $\mathbf{S}^\text{AB}$ are elements of the total Fock matrix and overlap matrix described over the basis functions of subsystems A and B.  These equations are then analogously defined for the Fock matrix of B in A.  
Upon freeze-and-thaw convergence at the DFT level, the $\mathbf{h}^\text{A-in-B}[\gamma^\text{A}, \gamma^\text{B}]$ is used as the one-electron Hamiltonian for the \iFCIQMC calculation; thus, embedding only influences the one-electron integrals for the \iFCIQMC calculation. The final embedding energy is then
\begin{equation}
E_{\mathrm{total}} = E_{\mathrm{KS-DFT}}^{\mathrm{total}} -
E_{\mathrm{DFT-in-DFT}}^{\text{A}}
+ E_{\mathrm{iFCIQMC-in-DFT}}^{\text{A}},
\end{equation}
where $E_{\mathrm{KS-DFT}}^{\mathrm{total}}$ is the full-system Kohn-Sham  (KS)-DFT energy, $E_{\mathrm{DFT-in-DFT}}^{\text{A}}$ is the DFT energy of subsystem A embedded in the DFT potential of the rest of the system, and $E_{\mathrm{iFCIQMC-in-DFT}}^{\text{A}}$ is the \iFCIQMC energy of subsystem A embedded in the DFT potential of the rest of the system.

\subsection{Calculation details} %

The atomic coordinates of the systems under investigation were generated using the dispersion-corrected M06-D3 functional and the aug-cc-pVTZ basis set as implemented in Gaussian16. The geometries are presented in the Supplementary Information. %
Six frozen orbitals were used for the \bz{LiH} canonical systems, and seven frozen orbitals were used for the \bz{HF} canonical systems. No frozen orbitals were used for the embedded integrals of either system, nor the isolated diatomics. %
In our implementation, QSoME was modified to output integrals for \iFCIQMC using PySCF, \cite{sun_pyscf:_2018} which were then read into the HANDE software package.\cite{spencer_hande-qmc_2019}
For the dissociation curve of \ce{HF},MOLPRO \cite{MOLPRO-WIREs} was also used taking advantage of an already-existing interface with QSoME. 
These integrals consisted of single-particle Hartree--Fock eigenvalues ($\epsilon_i$) and electron repulsion integrals ($v_{ijkl}$).

The \iFCIQMC calculations were performed using the open-source code HANDE-QMC. For the \bz{LiH} system, an imaginary time step of  $2 \times {10^{-6}}$ a.u. was used with 200,000 reports and 20 Monte Carlo cycles between reports. For the \bz{HF} and \bz{F^-} systems, a smaller time step of $9 \times {10^{-7}}$ a.u. was used due to the additional electrons present, with 400,000 reports for the first three target populations and 600,000 reports for the largest three target populations. A larger time step of 0.002 a.u. was used for the isolated LiH, HF and the embedded systems, except for the 5 and 6 \AA\ separations, which used a timestep of 0.0002. In order to converge the calculations with respect to the target population, a range of target populations between ${10^{1}}$ and ${10^{6}}$ was used.

Without the embedding algorithm, the LiH physisorbed on benzene system contains 34 electrons, requires $2.8 \times 10^{41}$ determinants, and has a storage cost of 700 MB. After embedding is introduced, the subsystem treated with \iFCIQMC is reduced to 4 electrons and $2.9 \times 10^{4}$ determinants, with an integral storage cost of 440 KB. 

\section{Results and discussion} %

It is common for energy \emph{differences} to yield better convergence (with respect to excitation rank, for example, in coupled cluster theory) than total energies themselves; this phenomenon, known as error cancellation, is a common benefit of running quantum-chemical calculations.
In \iFCIQMC (in common with FCIQMC), a walker population of a given size ($N_w$) represents the wavefunction. The calculation is only exact if it is converged with respect to this walker number. 
An under-explored issue of \iFCIQMC calculations is that convergence is not faster for energy differences than for individual energies. 
The dissociation energies of LiH on benzene and HF on benzene represent two paradigmatic examples of how dissociation energies can be extremely challenging and costly to converge in \iFCIQMC due to a lack of error cancellation between reactants and products.

We hypothesize that adding benzene to straightforward LiH and HF dissociation energy calculations will cause \iFCIQMC to fail in a way that can be remedied by using embedding. To test our hypothesis, we calculate the energy changes associated with four reactions: 
\begin{equation}
\begin{split}
\ce{LiH &-> Li+ + H-} \\
\ce{C6H6\bond{-}LiH &-> C6H6\bond{-}Li+ + H-} \\
\ce{HF &-> H+ + F-} \\
\ce{C6H6\bond{-} HF &-> C6H6\bond{-}F- + H+} 
\end{split}
\end{equation}
Here, we are required to use the closed-shell ionic dissociation products by the embedding code; an open shell implementation is planned.
We note that in \bz{HF}, the H atom is closest to the benzene ring and, following \ce{H+} removal and geometry optimization, the \ce{F-} migrates into the plane of the ring.  
In particular, we reason that the dissociation energy of a LiH or HF molecule physisorbed to benzene will be significantly more difficult to calculate using \iFCIQMC due to non-monotonic energy convergence with system size $N$. 
In contrast with other methods, \iFCIQMC does not show error cancellation between systems that contain different numbers of electrons. 

\begin{figure}
\begin{center}
\subfigure[\mbox{}]{%
\includegraphics[width=0.4\textwidth,height=\textheight,keepaspectratio]{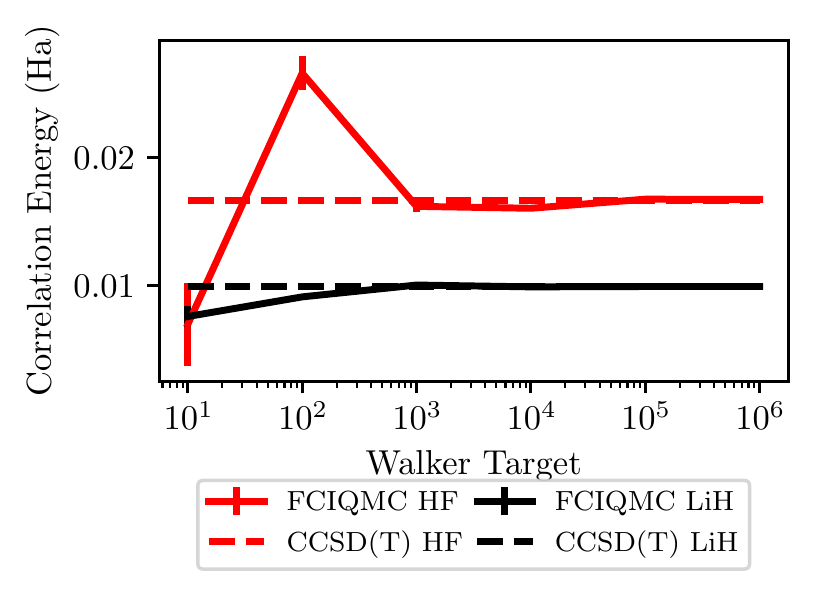}
\label{IsolNewLabel}
}

\subfigure[\mbox{}]{%
\includegraphics[width=0.4\textwidth,height=\textheight,keepaspectratio]{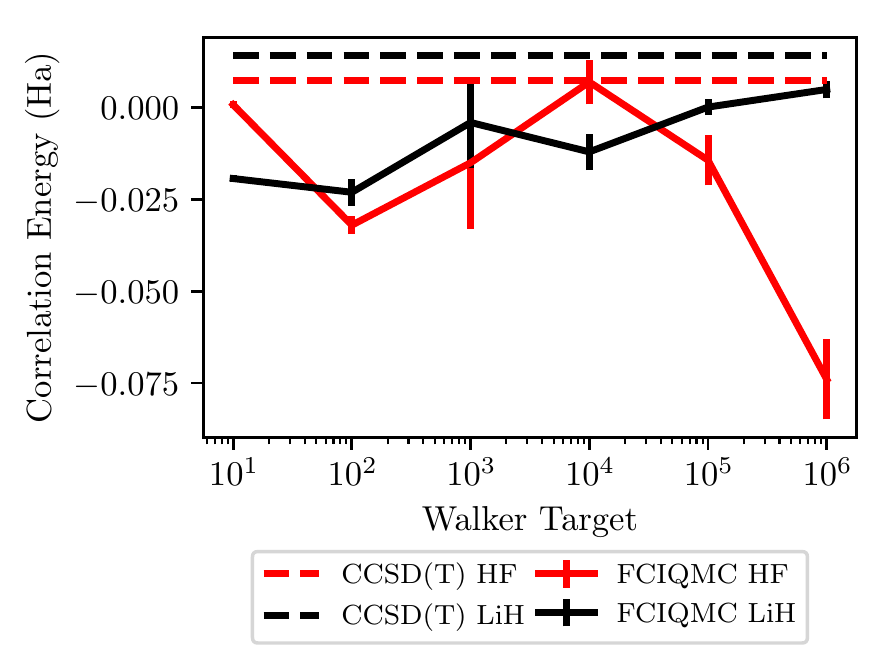}
\label{FullNewLabel}
}

\subfigure[\mbox{}]{%
\includegraphics[width=0.4\textwidth,height=\textheight,keepaspectratio]{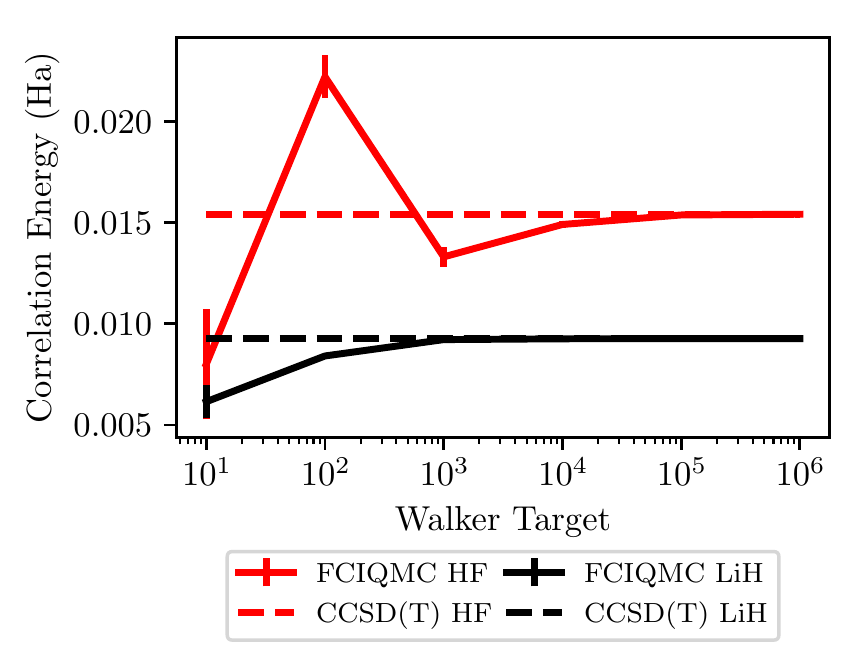}
\label{Embed}
}
\caption{
Correlation energy contribution to the dissociation energies of cc-pVDZ \ce{LiH} and \ce{HF} for molecules that are (a) isolated  (4 and 10 electrons respectively), (b) physisorbed to benzene (34 and 38 electrons respectively), (c) physisorbed to benzene and embedded (4 and 10 electrons treated explicitly with \iFCIQMCbracket). %
The \iFCIQMC calculations, shown as solid lines, were performed with six target populations ranging from $10^1$ to $10^6$ on a logarithmic scale. Good agreement is achieved between \iFCIQMC and CCSD(T) for isolated and embedded systems. }
\label{Rxn Energies 1}
\end{center}
\end{figure}

Figure \ref{Rxn Energies 1} shows data we collected in support of our claim. This data is also presented in table form in the Supplementary Information. 
Each of these plots is an initiator convergence plot, where the walker population is varied from $10^1$ to $10^6$, and the energy is computed using \iFCIQMC. 
We plot the \iFCIQMC energy differences between reactants and products for the LiH and HF dissociation reactions, and compare these differences to CCSD(T) dissociation energies. 
CCSD(T) can serve as a good benchmark for initiator convergence: initiator error can vary greatly over many orders of magnitude in energy, and CCSD(T) is generally thought to have systematic error only on the order of 1 millihartree.

Figure \ref{IsolNewLabel} shows that isolated \ce{LiH} and \ce{HF} dissociation energies rapidly converge as a function of walker number, showing complete convergence at $10^4$ and $10^5$ walkers, respectively. The \iFCIQMC and CCSD(T) results are in agreement with each other to within 1 millihartree for $N_w \geq 10^3$, and within 10 millihartree for the smaller target populations. 
The \ce{HF} dissociation converges in an oscillatory manner, because \ce{HF} is slightly slower to converge than \ce{F-}; in general, fine-grained oscillatory convergence has been shown in individual calculations. \cite{shepherd_investigation_2012} 
The \ce{HF} system contains more variability at lower walker numbers than the \ce{LiH} system, as is expected due to the higher number of electrons present in \ce{HF}.  
As we expect, our results show that the isolated systems with small numbers of electrons converge with only modest convergence errors.

In contrast to the isolated molecules, convergence is difficult for the dissociation of molecules physisorbed on benzene.  
The convergence difficulties for these systems are shown in \reffig{FullNewLabel}, where the oscillatory behavior observed in \reffig{IsolNewLabel} is magnified; in the case of HF, we are not able to converge this calculation at all in order to obtain a reaction energy, as the energy difference between $10^5$ and $10^6$ walkers is approximately -0.0597 hartree.  
Physisorption onto benzene adds 30 electrons to the isolated molecules; thus, significantly harder convergence is unsurprising.
Again, since \iFCIQMC does not show error cancellation between systems containing different numbers of electrons, \ce{C6H6\bond{-} HF} and \ce{C6H6\bond{-} F-} converge at different rates, which causes the energy difference between these two systems to be oscillatory. 
This is a key result of this manuscript that we explore later in further detail.

In \reffig{Embed}, we present the results of the \iFCIQMCbracket-in-DFT embedded systems.  Since embedding decreases the number of electrons treated directly by \iFCIQMCbracket, we are able to converge the \iFCIQMC energies of \ce{C6H6\bond{-} LiH} and \ce{C6H6\bond{-} HF} as easily as isolated LiH and HF.
We see similar oscillatory behavior in the embedded calculations as we do for the isolated systems: 
Target populations $10^1$ and $10^2$ are still not very accurate.  Fortunately, as we increase the target population, we see clear convergence.
Comparing the three initiator curves across \reffig{Rxn Energies 1} reveals a similar convergence trend. This is a very encouraging result, as it shows the \iFCIQMCbracket-in-DFT embedding gives convergent results while simultaneously reducing the cost of these calculations significantly. 

As computational cost is proportional to walker number, the ability to converge a calculation at $10^3$ walkers compared with leaving it unconverged at $10^6$ walkers represents a cost savings of at least 1000x. Data we present in the SI additionally show a 1000x savings in memory. 

We fully appreciate that there is an unquantified embedding error in these calculations. 
This causes a change in ordering of the \bz{HF} and \bz{LiH} dissociation energies between \reffig{FullNewLabel} and \reffig{Embed} at the CCSD(T) level.
For completeness, we note that the difference between CCSD(T) embedded calculations and full-system calculations give us an estimate of the \iFCIQMC embedding error as 4.31 millihartree and 8.01 millihartree for LiH and HF, respectively.  However, our previous studies have shown that the embedding error can be further decreased by enlarging the wavefunction subsystem.\cite{Chulhai2018Projection-BasedSystems}
Although we are interested in quantifying the \iFCIQMC embedding error and using it to benchmark embedded CCSD(T), this analysis is beyond the scope of the proof-of-principle offered by this paper. We now analyze the sources of error and the way that embedding overcomes convergence difficulties in \iFCIQMCbracket.

\subsection{Analysis of different convergence behaviors in \iFCIQMC} %

\begin{figure}
\begin{center}
\subfigure[\mbox{}]{%

\includegraphics[width=0.4\textwidth,height=\textheight,keepaspectratio]{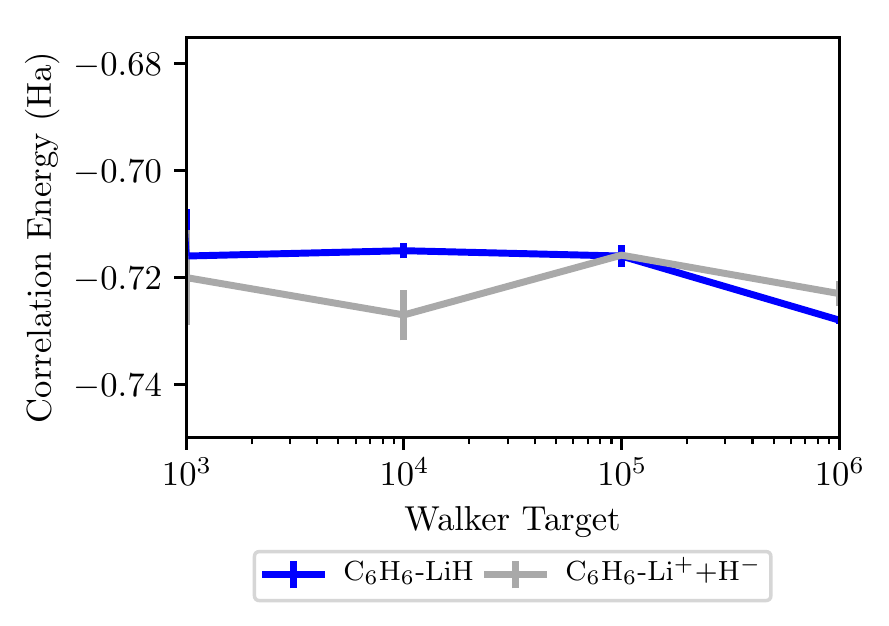}}
\label{LiHinit}

\subfigure[\mbox{}]{%
\includegraphics[width=0.4\textwidth,height=\textheight,keepaspectratio]{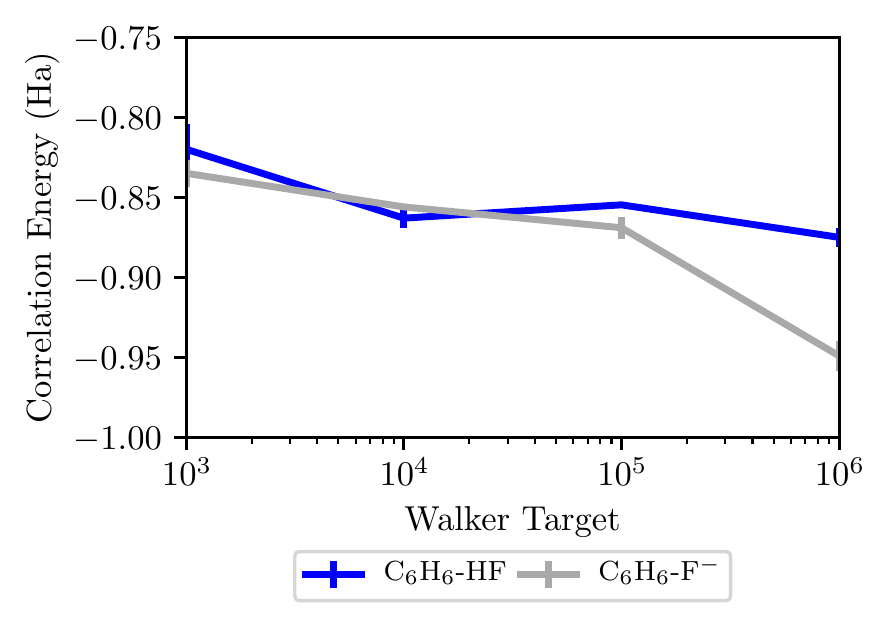}}
\label{HFinit}

\caption{The initiator curves at walker numbers $N_w = 10^3$ through $10^6$ for the products and reactants of the dissociation reactions of (a) LiH and (b) HF physisorbed on benzene.
 } 
\label{InitiatorConvergence}
\end{center}
\end{figure}

There are a number of analyses we can conduct in order to probe the extent of the non-convergent behavior described above in \reffig{FullNewLabel}---the case where all electrons in the benzene molecule are fully present in the \iFCIQMC calculation.
In \reffig{InitiatorConvergence}, the convergence of the reactants and products of dissociation for \bz{LiH} and \bz{HF} are shown. 
It can be seen from this figure that these calculations are not converged with respect to the number of walkers. This represents a particularly severe case where reactant and product energies actually cross over, which causes the energy differences to oscillate rather than converge smoothly, as observed in \reffig{FullNewLabel}.

Both \bz{LiH} and \bz{HF} represent different types of challenges in convergence.
In \bz{HF}, where reactants and products have the same number of electrons, each \iFCIQMC calculation appears to be smoothly converging as a function of walker number. 
Prior work has established the appearance of such smooth convergence as a stretched exponential in the walker population,  $\exp(-N_w^\alpha), \alpha << 1.0$.\cite{booth_breaking_2011}
The decay parameters are highly system-dependent, and as such, two converging calculations could easily cross over one another. 
The general form of two converging calculations is:
\begin{equation}
E_\mathrm{corr, A-B}=E_{corr,A}-E_{corr,B}+A_1 e^{-N_w^{\alpha_1}}-A_2 e^{-N_w^{\alpha_2}}
\end{equation}
In the case of HF, the combined initiator error, $A_1 e^{-N_w^{\alpha_1}}-A_2 e^{-N_w^{\alpha_2}}$, obscures or is much larger than the term $E_{corr,A}-E_{corr,B}$.  As a result, the reaction energy fails to converge, instead oscillating even at large walker numbers.

The underlying reason for the differences in convergence between \bz{HF} and \bz{F-} is not known.
It seems likely that the form of the stretched exponential is itself related to excited state decays in imaginary time, although this has not been established in the literature. 
Specifically, the overlap between the simulation wavefunction in imaginary time, $|\Psi (\tau) \rangle$, and the FCI excited states, $| \Psi_i \rangle$, is expected to decay exponentially in imaginary time:\cite{booth_breaking_2011}
\begin{equation}
\langle \Psi (\tau) | \Psi_i \rangle = C_i \exp (- \tau (E_i-E_0) )
\end{equation}
where $E_i$ and $E_0$ are the excited state and ground state energy eigenvalues, respectively.
In this picture, then, a simulation with insufficient walker population would have to get stuck somewhere between one state and another in a way that cannot be resolved by projecting out over more imaginary time steps, because there is not enough information in each timestep to afford resolution of the ground state. 

The case of \bz{LiH} is a little different, since \bz{Li+} exhibits oscillatory convergence already. This case of oscillatory fine structure has been seen before, such as in studies of the uniform electron gas. \cite{shepherd_investigation_2012} 
This on its own hampers convergence, lending an oscillatory character to the reaction energy independent of whether these calculations are themselves converging to the correct energy.

\subsection{Hartree--Fock Population} %

\begin{figure}
\includegraphics[width=0.4\textwidth,height=\textheight,keepaspectratio]{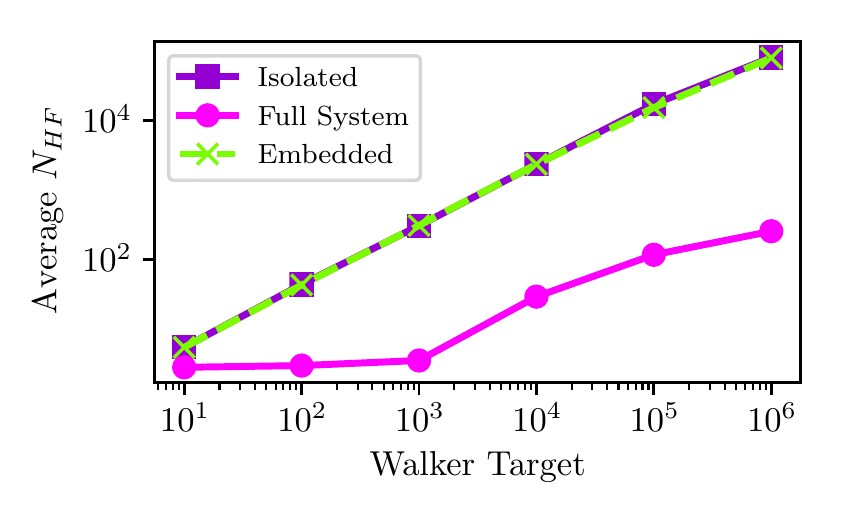}
\caption{The population of walkers on the Hartree--Fock determinant in the \iFCIQMC calculation with respect to iteration for each target population of $N_w =10^1$ to $10^6$ for each of three LiH systems: isolated LiH, the full system \bz{LiH} and the embedded \bz{LiH}.} 
\label{HF_populations}

\end{figure}

Another measure by which we can compare the isolated and embedded calculations is the number of walkers present on the Hartree--Fock determinant (shown in \reffig{HF_populations}). 
This population is sometimes used as a means to determine convergence of an \iFCIQMC calculation, since, in the early phase of an \iFCIQMC calculation, it does not vary from its baseline of $[\mathcal{O}(1)]$ walker. 
The number of walkers on the Hartree--Fock determinant also confirms the different convergence behaviors of the full, isolated, and embedded systems:
The embedded and isolated systems have Hartree--Fock populations that grow at the same rate, whereas the full system has many less of these kind of walkers.
In terms of the walker dynamics, the larger number of determinants in the full system depletes the signal present on the Hartree--Fock determinant and slows convergence.

\subsection{Embedding and the sign problem in \iFCIQMC} %

\begin{figure}
\begin{center}
\subfigure[\mbox{}]{%
\includegraphics[width=0.4\textwidth,height=\textheight,keepaspectratio]{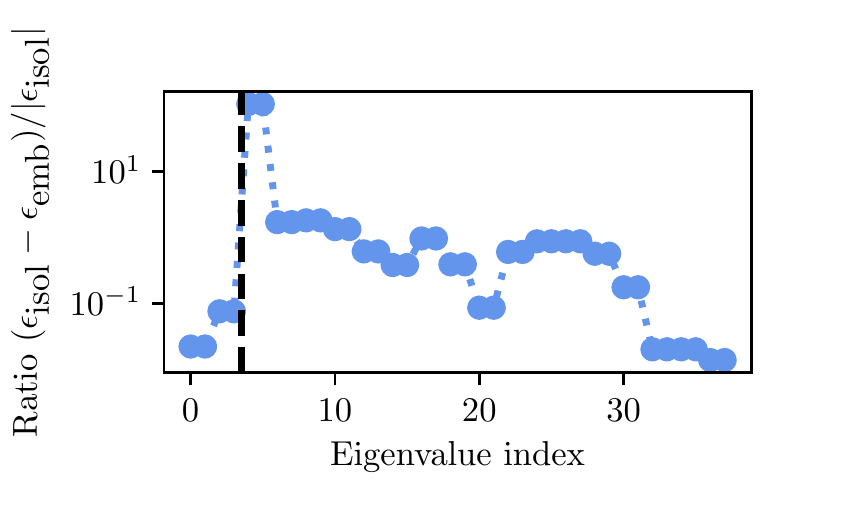}
\label{eigenratios}
}

\subfigure[\mbox{}]{%
\includegraphics[width=0.4\textwidth,height=\textheight,keepaspectratio]{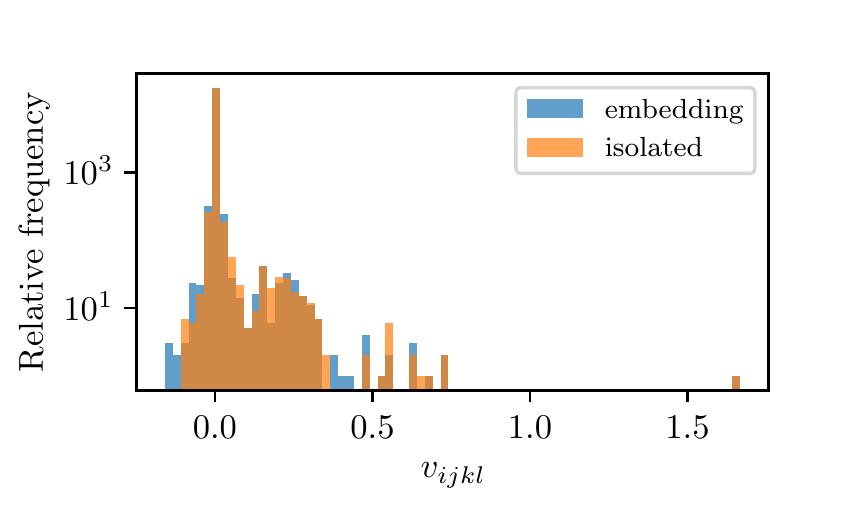}
\label{ERIfig}
}
\caption{Changes in the \ce{LiH} integral table for \iFCIQMC represented through (a) differences between eigenvalues $\epsilon_{i}$ for the embedding and isolated systems, where the black dashed line represents the division between occupied and virtual Hartree--Fock orbitals, and (b) electron repulsion integrals $v_{ijkl}$ for both embedded and isolated systems.}
\label{integral_comparison}
\end{center}
\end{figure}

The sign problem in FCIQMC has been related to the amount of spin frustration in the system: each Slater determinant in the system needs to find its sign over the course of a simulation.\cite{spencer_sign_2012} Specifically, the eigenvalue of a matrix $H_{ij}^\prime= \delta_{ij} H_{ij} - (1-\delta_{ij}) | H_{ij}|$, where $\delta_{ij}$ is the Kronecker delta, whose eigenstate has entirely non-negative components and contaminates solutions. 

The signs in $H$ come from the four-index integrals via the Slater--Condon rules, and so it is important to discuss whether there is a significant change in the integrals due to embedding. 
In \reffig{integral_comparison}, we show a comparison of two types of integrals that are passed between the embedding code and \iFCIQMC for isolated \ce{LiH} compared with embedded \bz{LiH}.
In this case, the LiH eigenvalues are generally lowered by between -0.01 hartree to -0.3 hartree by embedding.
The specific ratio for each eigenvalue is plotted against its energy-ordered index in \reffig{eigenratios}, showing that as the eigenvalue becomes higher in energy, it is also affected less by embedding. 
We show the effect on the electron repulsion integrals in \reffig{ERIfig}, where the distribution of the $\sim1800$ integrals is presented as a histogram. %
The molecular orbitals differ between the isolated case and the embedding case and this leads to the small changes in the electron repulsion integrals. 
From the plots above, we would expect that there is not an increase in the complexity of the sign problem, since most matrix elements remain unchanged. 

\subsection{Application to bond stretching}

\begin{figure}
\begin{center}
\subfigure[\mbox{}]{%
\includegraphics[width=0.4\textwidth,height=\textheight,keepaspectratio]{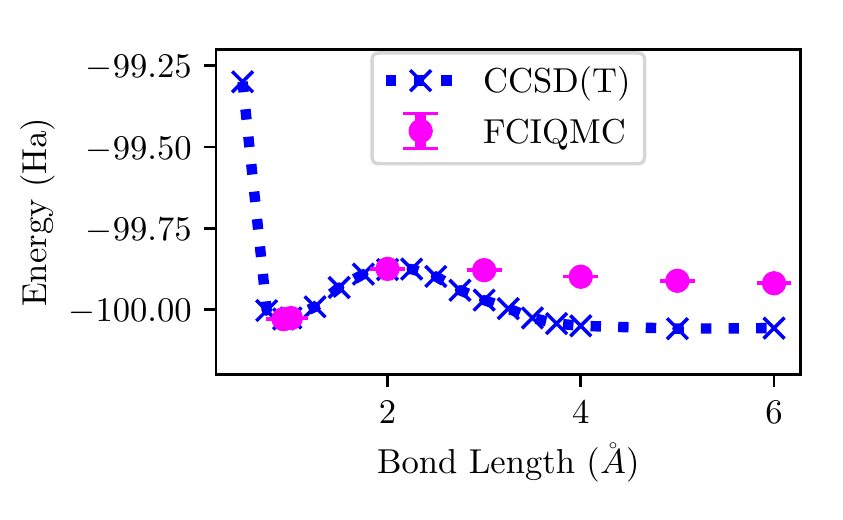}
\label{new_fig_a}
}

\subfigure[\mbox{}]{%
\includegraphics[width=0.4\textwidth,height=\textheight,keepaspectratio]{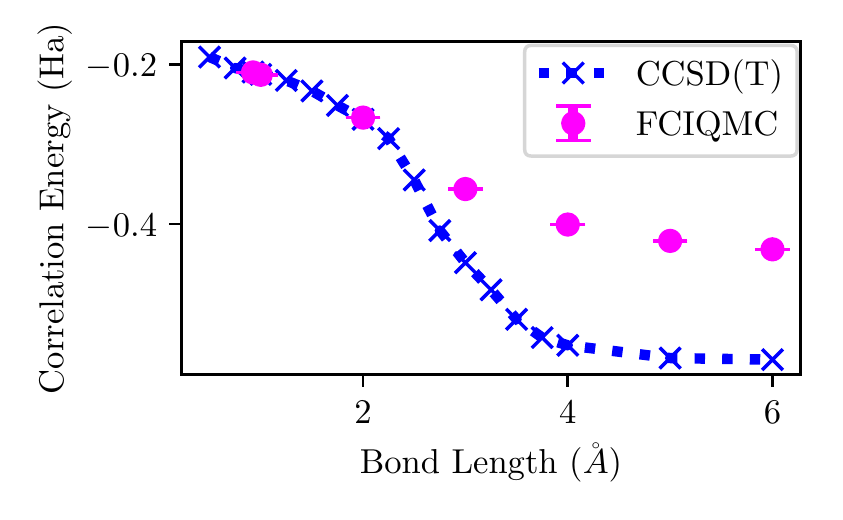}
\label{new_fig_b}
}
\caption{Bond dissocation energy curves for cc-pVDZ hydrogen fluoride molecule embedded on benzene, showing \iFCIQMC has improved accuracy over CCSD(T) for (a) total energies and (b) correlation energies. 
These graphs show agreement between the two methods between the equilibrium separation and 2.00 \AA, but the two methods diverge at longer separations.
CCSD(T) calculations are shown as blue dashed lines and \iFCIQMC calculations are shown as fuchsia circles. 
The CCSD(T) calculations were performed on atomic separations from 0.50 \AA\ to 4.00 \AA\  in 0.25 \AA\ increments, as well as the equilibrium separation of 0.92 \AA\, and separations of 5 and 6 \AA. \iFCIQMC calculations were added at the equilbrium geometry, 1.00, 2.00, 3.00, 4.00, 5.00 and 6.00 \AA~separation. 
  } %
\label{new_fig}
\end{center}
\end{figure}

Bond dissociation energy curves are frequently used to benchmark new developments in FCIQMC~\cite{booth_breaking_2011}.
This is in part because CCSD(T) is known to fail due to the strong correlation which occurs as the bond is stretched, leading to certain determinants becoming closer in energy while being strongly coupled.\cite{bartlett_coupled-cluster_2007}
In order to highlight the potential benefits of FCIQMC to the study of catalysis we can therefore make comparison between FCIQMC and CCSD(T) for a bond dissociation curve.

Here, we model the dissociation of \ce{H-F} in \bz{HF} by increasing the \ce{H-F} bond distance. This represents the following dissociation:
\begin{equation}
\ce{C6H6\bond{-} HF -> C6H6\bond{-}H \bond{...} F} 
\end{equation}
where the dissociation products, by contrast to Eq. (11), show dissociation by drawing the \ce{F-} away from the molecule with the rest of the geometry frozen.

Figure \ref{new_fig} shows total and correlation energies calculated at different \ce{H-F} separations.
At above 2\AA, the CCSD(T) energy decreases in a manner indicative of strong correlation.
By contrast, FCIQMC energies appear to level off to an overall correlation contribution to the bond dissociation energy of approximately 
$-0.221(1)$ Ha (between the equilibrium separation and 6\AA).
We note in passing that this is different from the previous correlation contribution to bond dissociation energy of $0.0154(2)$ Ha %
because these fragments are not the same (see Eq. (14) and Eq. (11)).

\section{Conclusions}

In summary, we here examined convergence difficulties present when using \iFCIQMC to calculate the electronic structure of large systems by exploring the bond dissociations of two prototypical molecules, LiH and HF, physisorbed to benzene. 
Since \iFCIQMC does not show error cancellation between systems with different numbers of electrons, the energy differences between reactants and products tended to oscillate.
As a result, dissociation energies calculated from \iFCIQMC did not converge.  To remedy the convergence issues that \iFCIQMC has with large systems, we embedded \iFCIQMC in DFT.  We showed that this new embedded \iFCIQMC was better able to converge dissociation energies, giving results that agree with our CCSD(T) benchmarks.  
By way of an application, we have also shown the ability of \iFCIQMC-in-DFT to more accurately model dissociation curves at atomic separations greater than equilibrium than CCSD(T). 
This demonstrates the ability of the absolute localization approach for Huzinaga projection-based embedding to treat strongly correlated systems using high-level \iFCIQMC wavefunctions embedded in DFT. 

Since embedded \iFCIQMC also reduces the number of electrons (and thus orbitals) in a calculation, embedded \iFCIQMC calculations run with substantially lower cost than full \iFCIQMCbracket,  alleviating the method's reduced-exponential cost scaling. Based on our results, we estimate the cost saving to be at least 1000x in compute time and 1000x in memory for the model systems studied here.  Whereas for larger systems, \iFCIQMC calculations can be computationally intractable while embedded \iFCIQMC calculations will remain feasible.   There are applications for which CCSD(T) fails to give good answers, such as those involving strong correlation or bond breaking; \iFCIQMC can treat these applications with high accuracy. As such, we believe that \iFCIQMC emdedded in DFT is a significant and realistic step forward for bringing \iFCIQMC towards the routine treatment of real applications, as DFT embedding both alleviates convergence concerns and dramatically reduces the cost of the method. 

More broadly, the dissociation curve we calculated represents a situation where strong correlation (bond breaking) was treated by QMC and weak correlation (physisorption) was treated by embedding and DFT; this is likely the best-case scenario for our method. To extend this work, we would move towards real systems.  A similar embedding approach as the one we take here has already shown promise for being applied to catalysis.\cite{lee_projection-based_2019} %
For example, quantum embedding was applied to a variety of Co-based catalysts to explore the coupling of the electronic structure of the transition metal to that of the ligand in the hydrogen evolution reaction,\cite{huo_breaking_2016} %
and to explore the multireference character of these systems that presents challenges for DFT when calculating reaction barriers. \cite{welborn_even-handed_2018} 
Also, Carter and coworkers have used a similar but distinct embedding method to study \ce{H2} dissociation on Au and Al nanoparticles.\cite{mukherjee_hot_2013, zhou_aluminum_2016}  In our example of HF dissociation, there was no barrier to dissociation, however in the general case a bond dissociation curve could be used to determine transition state energies and therefore kinetic barrier heights.

We believe that our work is particularly timely because there has been a call from prominent researchers studying oxygen reduction catalysts \cite{kulkarni_understanding_2018} %
to focus on the understanding and design of multi-functional active sites for next-generation catalysts and suggested embedding methods could help us get there. In examples such as bifunctional sites or in confinement, we may well expect strong correlation which is quantum mechanically coupled to the environment, and we expect FCIQMC-in-DFT embedding to find applications there.

There are limitations to the embedding approach which are relevant for high-accuracy modelling. The general case where this method will work is when the density from DFT is almost exact. This is especially true for the DFT region, whose density does not change in this method due to our using a frozen density approach. In practice, there are studies that have explored the severity of this approximation\cite{goodpaster_accurate_2014} and found that much can be gained by allowing the errors in the density outside the embedded (here, QMC) region to cancel. It is still very much an open question as to whether a region including strong correlation could be left in the embedded region. 

Since strong correlation is often investigated by way of model systems, we note the conditions to apply this approach to model systems as follows. 
The model system would need to have identifiable localized fragments (here, atoms) and orbitals that are associated with that local fragment which have a dot product rule. 
The formalism by which the subsystems are divided is exact within a Kohn--Sham formalism and partitions the subsystems to have integer numbers of electrons. In principle it would be possible to use non-integer subsystems which has been applied to 1-D hydrogen chain systems.\cite{elliott_partition_2010, tang_fragment_2012} %

One other limitation of this work is that we have not analyzed the added error in the correlation energy introduced when undertaking embedding, since we believe it is outside of the scope of a proof-of-principle and deserves much more attention on its own.
Since CCSD(T) can treat the full systems for the prototypical bond dissociations studied in this manuscript, we \emph{could} have added a correction to our embedded \iFCIQMC arising from the CCSD(T) energy difference between the full and embedded systems; this may be a way forward for future work. 
It is also of note that the embedding error has been analyzed for CCSD-in-DFT in comparison to CCSD\cite{Chulhai2017} and also for CCSD(T)-in-DFT in comparison with experiment;\cite{Chulhai2018Projection-BasedSystems} we would expect comparable errors at this level of theory. 
We could also treat a system that is small enough to examine the full system with \iFCIQMC, resulting in our being able to benchmark the embedding error for the benefit of other practitioners.

Further work will be forthcoming where this embedding is further developed for excited states and EA/IP calculations; FCIQMC can also be interfaces with other types of calculations such as those with periodic boundary conditions for which a separate periodic code exists. 
Benchmarking \iFCIQMC in comparison to other high-accuracy methods (CCSD(T), DMRG, selected CI) to find the relative advantages and disadvantages of each method represents a very interesting open question. To facilitate this, integral files and output files can be found at https://doi.org/10.25820/data.001111.

In closing, we believe that this study highlights an important step forward for both \iFCIQMC and embedding.  We believe that the work presented here brings the community one step closer to the routine application of high-accuracy electronic structure to study strongly-correlated systems of chemical and technological interest.  %

\section{Acknowledgements}
This research is supported by the Nanoporous Materials Genome Center, funded by the U.S. Department of Energy, Office of Basic Energy Sciences, Division of Chemical Sciences, Geosciences and Biosciences under Award DE-FG02-17ER16362.
DSG and JG acknowledge an award of computer time was provided by the ASCR Leadership Computing Challenge (ALCC) program. This research used resources of the National Energy Research Scientific Computing Center (NERSC), a U.S. Department of Energy Office of Science User Facility operated under Contract No. DE-AC02-05CH11231. DSG and JG acknowledge additional computer resources were provided by the Minnesota Supercomputing Institute (MSI) at the University of Minnesota.
JJS and HRP acknowledge the University of Iowa for funding and the University of Iowa Informatics Initiative (UI3) for computer resources. The code used throughout this work was HANDE (hande.org.uk), QSoME (github.com/Goodpaster/QSoME), and PySCF (sunqm.github.io/pyscf/).

\providecommand{\latin}[1]{#1}
\makeatletter
\providecommand{\doi}
  {\begingroup\let\do\@makeother\dospecials
  \catcode`\{=1 \catcode`\}=2 \doi@aux}
\providecommand{\doi@aux}[1]{\endgroup\texttt{#1}}
\makeatother
\providecommand*\mcitethebibliography{\thebibliography}
\csname @ifundefined\endcsname{endmcitethebibliography}
  {\let\endmcitethebibliography\endthebibliography}{}

 \end{document}